\documentclass[letter]{article}

\usepackage{amsmath,amssymb}
\usepackage{graphicx}
\usepackage{dcolumn}
\usepackage{bm}

\usepackage[affil-it]{authblk}

\date{}
\begin{document}

\title{Negative-resistance models for parametrically flux-pumped superconducting quantum interference devices}

\author{K.M. Sundqvist\thanks{Electronic address: \texttt{kyle.sundqvist@gmail.com}; Corresponding author} }

\author{P. Delsing}

\affil{Microtechnology and Nanoscience,\\ Chalmers University of Technology,\\ SE-412 96 G\"{o}teborg, Sweden }

\maketitle

\begin{abstract} 
A Superconducting QUantum Interference Device (SQUID) modulated by a fast oscillating magnetic flux can be used as a parametric amplifier, providing gain with very little added noise. Here, we develop linearized models to describe the parametrically flux-pumped SQUID in terms of an impedance. An unpumped SQUID acts as an inductance, the Josephson inductance, whereas a flux-pumped SQUID develops an additional, parallel element which we have coined the ``pumpistor."  Parametric gain can be understood as a result of a negative resistance of the pumpistor. In the degenerate case, the gain is sensitive to the relative phase between the pump and signal.  In the nondegenerate case, gain is independent of this phase.

We develop our models first for degenerate parametric pumping in the three-wave and four-wave cases, where the pump frequency is either twice or equal to the signal frequency,  respectively.  We then derive expressions for the nondegenerate case where the pump frequency is not a multiple of the signal frequency, where it becomes necessary to consider idler tones which develop. For the nondegenerate three-wave case, we present an intuitive picture for a parametric amplifier containing a flux-pumped SQUID where current at the signal frequency depends upon the load impedance at an idler frequency.  This understanding provides insight and readily testable predictions of circuits containing flux-pumped SQUIDs.

\end{abstract}




\section{Introduction}

Parametric amplifiers are attractive in that they can in principle amplify a signal while only adding a minimum of noise\cite{Caves:1982}. From this point of view, parametric amplifiers may be divided into two groups; \emph{phase sensitive} amplifiers which amplify only one of the incoming quadratures, and \emph{phase insensitive} amplifiers which amplify both quadratures, thereby preserving the phase of the signal.  A phase sensitive amplifier can in principle amplify the signal without adding any noise. The minimum noise added by a phase insensitive amplifier corresponds to half a quantum of the amplified frequency, $\hbar \omega /2$.

In a parametric amplifier, some parameter of the system must be varied in time. By pumping the system, \textit{i.e.} modulating that parameter at one frequency, it is possible to amplify a signal at a different frequency. Power is transferred from the pump frequency to the signal frequency.

Parametric amplifiers can be realized in a large number of systems, both in optics and in electronics. A typical example in optics is a fiber-based amplifier where the refractive index of the fiber material is modulated by the pump.  In other systems utilizing varactor diodes, it is the nonlinear diode capacitance which is modulated.  Varactor diodes are typically used at frequencies ranging from radio to THz frequencies. 

Superconducting circuits can also be used to build parametric amplifiers in the microwave domain. The use of parametric amplifiers with Josephson junctions was pioneered by several researchers in the 1970's\cite{Feldman:1975, Taur:1977, Feldman:1977,Wahlsten:1977,Wahlsten:1978}, as well as Bernard Yurke in the 1980's\cite{Yurke:1988,Yurke:1989}.  Josephson junctions are used as parametric inductances, and may be pumped either by a time varying current through the junction\cite{CastellanosBeltran:2007,Eichler:2011,Hatridge:2011}, or in a SQUID geometry by a time-varying magnetic flux\cite{Yamamoto:2008, Hatridge:2011, Wilson:2012, Sundqvist:2013}. Alternatively the kinetic inductance of a thin superconductor can be used as the parametric component\cite{Tholen:2007,HoEom:2012}. 

Parametric amplifiers based on superconducting devices have recently regained interest because of the need for better amplifiers for qubit readout and microwave quantum optics.  The utility  of these amplifiers have been demonstrated in a number of experiments showing single shot qubit readout\cite{Mallet:2009}, quantum feedback\cite{Vijay:2012}, vacuum squeezing\cite{Flurin:2012}, and in determining the statistics of nonclassical photon states\cite{Steffen:2013}. There are two major advantages of superconducting parametric amplifiers: i) they have very low dissipation, and ii) they have well characterized and engineer-able properties.  This makes it possible to design well functioning parametric amplifiers with good gain and little added noise\cite{CastellanosBeltran:2007,Abdo:2011}.

To understand and implement a parametric amplifier, one often needs to solve a system of coupled equations where it may be difficult to fully appreciate the amplifier's overall properties.  Along with the resurgent use of parametric amplifiers as applied to quantum systems, a quantum optics formalism is also typically adopted to explain the amplifier.  

By contrast, we recently presented a linearized impedance model for a flux-pumped SQUID following the engineering formalism\cite{Blackwell:1961,Decroly:1973,Howson:1970} developed for (classical) varactor diodes in the 1960's and 70's.  While a similar formalism had also been utilized for early treatments of Josephson junction parametric amplifiers\cite{Feldman:1977}, this had not been applied to the flux-pumped SQUID. The flux-pumped SQUID can be represented as a parallel combination of a Josephson inductance and an additional circuit element which we named the ``pumpistor"\cite{Sundqvist:2013}. The pumpistor has the frequency dependence of an inductance, but it is an inductance which is \emph{complex}.  The phase of this complex inductance (or impedance) depends on the phase angle of the pump relative to the signal.  By properly adjusting the pump, the pumpistor can act as a negative resistance.  Thus, it can provide gain in the circuit. In this recent paper, we treated only the three-wave degenerate case, \textit{i.e.} where the pump is applied at exactly twice the signal frequency.

In this work, we extend this pumpistor model.  We revisit the three-wave degenerate case to include higher-order saturation effects.  We also explore the four-wave degenerate case, which couples to the pump at higher order.  Perhaps most importantly, we also treat the \emph{nondegenerate} case, where the pump frequency is not a multiple of the signal frequency.  Here a matrix formalism provides for the exploration of many types of nondegenerate frequency mixing, which, in addition to gain as a negative resistance, also describes up- and down- conversion of a signal. 


\section{The current response of a simple dc SQUID}

In this section, we briefly review the relations between external magnetic flux, effective junction phase, and series current in a dc SQUID. In this work, we refer to a dc SQUID simply as a ``SQUID," and we consider it free of parasitic internal impedances. To begin, we first consider a single Josephson junction in order to introduce the Josephson relations due to the \emph{dc} and \emph{ac} Josephson effects\cite{Josephson:1962}.

\subsection{Current and voltage in a simple Josephson junction}

In a Josephson junction, the \emph{dc Josephson effect} denotes the relation between the phase difference $\phi_J$, \textit{i.e.} the difference in phase between the superconducting order parameters on either side of the junction, and the current $I$ which flows through the junction.  This is given by
\begin{equation}
I = I_{c} \sin(\phi)\,.
\label{eqn:DCJosephsonEffect}
\end{equation}
Here, $I_{c}$ is the critical current for this single Josephson junction, which is its maximum allowed super-current.  The \emph{ac Josephson effect} relates the \emph{time derivative} of the phase difference to the voltage, V, across the junction.
\begin{equation}
V = \left( {\frac{{\Phi _0 }} {{2\pi }}} \right) ~\frac{d \phi}{dt},
\label{eqn:ACJosephsonEffect}
\end{equation} 
where, $\Phi_0 \equiv h/(2e)$ is the \emph{superconducting flux quantum}. By taking the time derivative of Eq. \ref{eqn:DCJosephsonEffect} and combining it with Eq. \ref{eqn:ACJosephsonEffect}, we see that the Josephson junction acts like an inductor, $dI/dt=V/L_J$, with the \emph{Josephson inductance}
\begin{equation}
L_J=\frac{\Phi_0}{2\pi I_c \cos{\phi}}\,.
\label{eqn:Josephson_Inductance}
\end{equation}

\subsection{ Extending the Josephson relations to a SQUID}

Placing two Josephson junctions (``1'' and ``2'') in parallel, we form a SQUID, where the currents combine as a sum.  We adopt the sign conventions suggested in Ref.\,\cite{Zagoskin:2011}.
\begin{equation}
I = I_{c1} \sin \left( {\phi _{1} } \right) - I_{c2} \sin \left( {\phi _{2} } \right).
\label{eqn:SQUID_Current_unsimplified}
\end{equation}
Going around the loop and returning to the same point, the phase can only subtend multiples of 2$\pi$. We therefore find a quantization condition for the superconducting loop flux. We regard the phase differences to occur only at the two Josephson junctions, \textit{i.e.}, neglecting the inductance of the loop. Furthermore we assume that the two junctions are equal, $I_{c1}=I_{c2}=I_c/2$, and we define the SQUID phase to be $\phi=(\phi_1-\phi_2)/2$. Then we arrive at the SQUID current,
\begin{equation}
I = I_c \cos \left(\pi \frac{\Phi_{\rm{ext}}}{\Phi _0} \right)\sin \left( \phi  \right).
\label{eqn:SQUIDCurrent}		
\end{equation}
We see that the SQUID acts just like a Josephson junction, but with a critical current tunable by the external magnetic flux $\Phi_{\rm{ext}}$. Note that our choice of sign convention following Ref. \cite{Zagoskin:2011} eliminates the need for taking the \emph{absolute value} of the quantity $\cos(\pi~\Phi_{\rm{ext}}/\Phi_0)$ in Eq. \ref{eqn:SQUIDCurrent}. This is not the case in the definition commonly used in other very good and popular references (\textit{e.g.}, \cite{VanDuzer:1999, Tinkham:1996}).  In any case, for this work we consider only the situation where $| \Phi _{\rm{ext}}/\Phi_0 | < | 1/2 |$.  Here, the quantity corresponding to $\cos(\pi~\Phi_{\rm{ext}}/\Phi_0)$ is always positive regardless of convention.

Thus, we recover a device phenomenology similar to the single Josephson junction depicted in Eqs. \ref{eqn:DCJosephsonEffect} and \ref{eqn:ACJosephsonEffect}.  Specifically, the SQUID acts as a tunable inductance such that
\begin{equation}
L_J=\frac{\Phi_0}{2\pi I_c \cos \left(\pi \frac{\Phi_{\rm{ext}}}{\Phi _0} \right) \cos{\phi}}\,.
\label{eqn:JosephsonInductance}
\end{equation}

In this section, we have defined the system of a SQUID by current and voltage relations similar to a single Josephson junction.  We found the SQUID to be tunable by an externally applied magnetic flux.  Using this framework, in the next section we examine the SQUID circuit response to a magnetic flux, \emph{applied dynamically}.

\section{The signal impedance of a SQUID, subject to a dynamically pumped external magnetic flux}

We investigate how a SQUID responds as an impedance due to the presence of a periodic perturbation of the external magnetic flux.  To this end, we assume the external flux is of the following form.
\begin{align}
\label{eqn:FluxPerturbation}
\Phi_{\text{ext}} & = \Phi_{\rm{dc}} + \delta \Phi  
\end{align}
Here $\Phi_{\rm{dc}}$ is a static (``quiescent'') magnetic flux, and we use a time-dependent perturbation of the form $ \delta \Phi  = \Phi_{\text{ac}}~\text{cos}(\omega_3 t + \theta_{3})$.

For convenience of notation, we define these following normalized flux amplitudes.
\begin{align}
F = \pi \frac{{{\Phi _{{\rm{dc}}}}}}{{{\Phi _0}}}\\
\delta f = \pi \frac{{{\Phi _{{\rm{ac}}}}}}{{{\Phi _0}}}
\end{align}

\subsection{An aside regarding labels and conventions}

For clarity, we take the opportunity to introduce a handful of electromagnetic disturbances necessary to understand our system. These small-signal disturbances occur at different frequencies.  We follow the nomenclature for frequency terms as presented by Blackwell and Kotzebue\cite{Blackwell:1961}.

Regarding frequencies and how we label them, in this work we consider at most six frequencies due to possible mixing effects.  Foremost, we consider a ``signal'' exists at a frequency assigned to index 1.  For a parametric amplifier, the signal frequency serves as the frequency of both the input and output of the device. We shall the small-signal current at the signal frequency, for a given voltage at this same frequency.  This gives us a ``signal impedance'' upon which we base our subsequent reasoning.  Some driven ``pump'' disturbance occurs at a frequency of index 3.  This pump frequency corresponds to the frequency at which the SQUID is driven externally. The pumping of the SQUID provides for a nonlinear interaction to occur.  Another frequency we consider is the ``idler'' frequency.  An idler response comes about due to the nonlinear mixing between signal and pump.  In the general case, we need to provide for the possibility for the idler response to exist, even if it remains as an internal state variable (serving neither as an externally accessible input or output to the circuit).   Among the various topologies which allow frequency mixing, an idler tone occurs at a frequency that is some linear combination of the signal and pump frequencies.  In this work we delineate an idler as either a \emph{sum} or a \emph{difference} between signal and pump frequencies, for either the \emph{three-wave} or \emph{four-wave} case.   An underlying principle of the parametric amplifier is that (some portion of) the power absorbed at the pump frequency is transferred to signal and idler frequencies, allowing for an amplified response.   

\begin{figure}[h!] 
\centering
   \includegraphics[width=3.5 in]{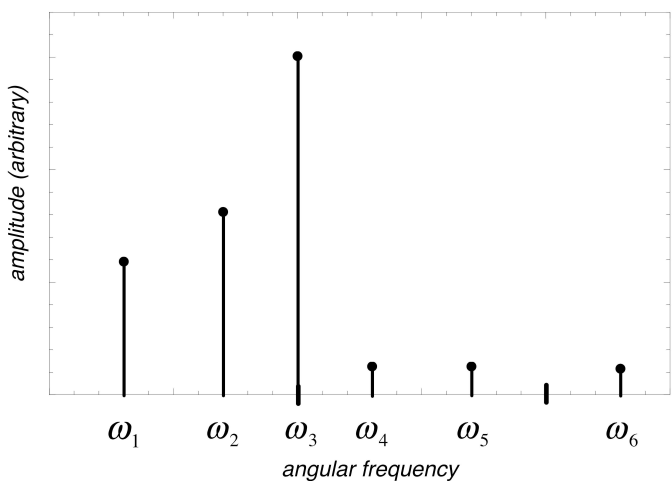} 
   \caption{This figure depicts the mixing terms we consider pertinent. The signal frequency is at (angular) frequency $\omega_1$, and the pump frequency is $\omega_3$. The amplitudes are arbitrary. }
   \label{Fig1}
\end{figure}

We list all considered mixing frequencies in Table \ref{tab:freq}, and provide a depiction in Fig. \ref{Fig1}.

\begin{table}[h!]
\caption{Our convention for the frequencies involved in mixing effects}
\label{tab:freq}
      \begin{tabular}{ccc}
        \hline
            (Angular) frequency  &  Designation   &  Relation \\ \hline
        ${\omega _1 }$  & ``signal'' & ${\omega _1 }$ \\
       ${\omega _2 }$ &  ``idler" (three-wave difference) &  ${\omega _3 } - {\omega _1 }$  \\
        ${\omega _3 }$ & ``pump"  &  ${\omega _3 }$  \\
         ${\omega _4 }$ & ``idler" (three-wave sum)  &   ${\omega _3 } + {\omega _1 }$  \\ 
         ${\omega _5 }$ & ``idler" (four-wave difference)  &   ${2 \omega _3 } - {\omega _1 }$  \\ 
          ${\omega _6 }$ & ``idler" (four-wave sum)  &   ${2 \omega _3 } + {\omega _1 }$  \\ \hline
      \end{tabular}
\end{table}

We must also consider different types of small-signal electromagnetic disturbances in the SQUID.  Generally, we may account for the \emph{junction phases}, \emph{voltages}, \emph{currents}, or \emph{magnetic fluxes}, at any of the signal, idler, or pump frequencies.  In practice, we account for any of the phases, voltages, and currents at the signal and idler frequencies, while only the magnetic flux at the pump frequency.  We assume periodic (cosine) forms for these quantities, including an offset phase term.  For instance, the phase across the junction at the signal frequency we denote as,
\begin{equation}
\phi_1(t) = {\tilde \phi_{1}} ~\text{cos}(\omega_1 t + \theta_{1}).
\end{equation}
To relate this phase to the corresponding voltage at the signal frequency, we use the Josephson relation of Eq. \ref{eqn:ACJosephsonEffect}.
\begin{equation}
v_1(t) = V_{1} ~\text{cos}(\omega_1 t + \theta_{v1})
\label{eqn:voltageForm}
\end{equation}
Above, we have set
\begin{equation}
V_{1} =  \frac{\Phi_0 ~{\tilde \phi_{1}} ~ \omega_1  }{2 \pi}, 
\end{equation}
as well as set the relation 
\begin{equation}
\theta_{v1} = \theta_{1} +  \pi/2.
\end{equation}

We have now established our nomenclature for small-signal responses in a flux-pumped SQUID. We proceed to treat the response of the SQUID under various, specific conditions.  We begin by studying the \emph{three-wave, degenerate} case.


\section{The three-wave degenerate case}
\label{sec:3waveDegen}

There are particular cases where we do not need to treat an idler response separate from the signal response.  For the three-wave case, at precisely half of the pump frequency, the signal and idler frequencies are identical.

\begin{center}
{\bf{The three-wave degenerate case:}}
\end{center}
\begin{align}
\omega_1  = \omega_2  = (\omega_3 - \omega_1)  = \omega_3 / 2
\label{eqn:3waveDegenerateFreq}
\end{align}

For this particular situation, we need only treat the signal and pump variables.  From Eq. \ref{eqn:SQUIDCurrent}, note that there are two terms to consider; the ``flux'' term, and the ``phase'' term.

\begin{equation}
i(t) = \underbrace {I_c \cos \left( {\pi ~\Phi _{{\text{ext}}}(t) /\Phi _0 } \right)}_{{\text{`flux'' term}}}\underbrace {\sin \left( \phi(t)  \right)}_{{\text{``phase'' term}}}
\label{SQUIDCurrentItemized}
\end{equation}

When treating these sorts of dynamics involving sinusoids, a common approximation is to implement Fourier-Bessel expansions \cite{VanDuzer:1999}. However, a simple Taylor expansion recovers the same result as a Fourier-Bessel expansion when approximating Bessel functions in their small-signal limit. We take separate series expansions of the two multiplied terms of Eq. \ref{SQUIDCurrentItemized}.  

To first order we expand the ``flux' term, using the flux-perturbation variable, $\delta \Phi$, of Eq. \ref{eqn:FluxPerturbation}.  We find

\begin{align}
   {I_c \cos \left( {\pi \Phi _{{\text{ext}}} /\Phi _0 } \right)} &  \approx   {I_c}\left[ {\cos \left( F \right) - \sin \left( F \right) \delta f  \cos \left( {{\omega _3}t + {\theta _3}} \right)} \right].
\label{eqn:IcCosExpand}
\end{align}

Also to first order, we expand the ``phase'' term so that sin$[\phi(t)] \approx \phi(t)$. Considering the functional form of the phase variables, we approximate the ``phase'' term by the following. 

\begin{equation}
\begin{gathered}
  \sin \left[ {\phi (t)} \right] \approx \phi (t) = \tilde \phi _{1} \cos \left( {\omega _1 t + \theta _{1} } \right)  \quad \quad \left( {{\text{degenerate case}}} \right)  \hfill \\
  \begin{array}{*{20}c}
   {} & {} & { = \tfrac{1}
{2}~\tilde \phi _{1}~ e^{j\theta _{p1} } e^{j\omega _1 t}  + }  \\

 \end{array} \tfrac{1}
{2}~\tilde \phi ^* _{1} ~e^{ - j\theta _{p1} } e^{ - j\omega _1 t}  \hfill \\ 
\end{gathered} 
\label{eqn:PhaseExpand}
\end{equation}
Here, since we have assumed a cosine dependence with an explicit phase angle, the amplitude $\tilde \phi_{1}$ is real and therefore equal to its complex conjugate $\tilde \phi^*_{1}$.  Yet, for now, we retain the use of conjugate notation for generality.  We also did not include a phase variable present at the pump frequency.  This is because we are interested in the signal response.  For frequency \emph{mixing} to occur, components at different frequencies must be \emph{multiplied}.  As long as the approximation $\sin[\phi(t) ] \approx \phi(t)$ is valid, the pump component of phase need not be included as it simply \emph{sums} with the signal component of phase.

The multiplication of Eq. \ref{eqn:IcCosExpand} by \ref{eqn:PhaseExpand} serves as our approximation to Eq. \ref{SQUIDCurrentItemized}.  We apply the degenerate frequency condition of Eq. \ref{eqn:3waveDegenerateFreq}. From the resulting expression, we find the terms present corresponding to the frequency component at $e^{ j \omega _{1} t }$.   We further consider the current to also be of a cosine response, 
\begin{align}
i_1(t)  = I_1 ~ \cos(\omega_1 t) = {\textstyle{1 \over 2}}\,~{I_1}~{e^{j{\omega _1}t}} + {\textstyle{1 \over 2}}\,~I^*_1~{e^{ - j{\omega _1}t}} = {i_1}{(t)_ + } + {i_1}{(t)_ - }
\end{align}
such that we can match its $e^{ j\omega _{1} t } $ component to the following form.
\begin{align}
{i_1}{(t)_ + }&{ = \frac{1}{2}~{I_1}~{e^{j{\omega _1}t}}}\\
{}&{ = \frac{1}{2}~{I_c}~{\tilde \phi _{1}}~\left[ {{e^{j{\theta _1}}}\cos \left( F \right) - \frac{{\delta f}}{2}\,\sin \left( F \right){e^{j\left( {{\theta _3} - {\theta _1}} \right)}}} \right]{e^{j{\omega _1}t}}}
\label{eqn:CurrentOmega1}
\end{align}
 
Considering a voltage based on the Josephson relation applied to the phase response, we find the following voltage component at $e^{ j\omega _{1} t }$.

\begin{align}
 \notag
 {v_1}{(t)_ + }&= \tfrac{1}
{2}\left( {V_{1} ~e^{j\theta _{v1} } } \right)e^{j\omega _1 t}  = \frac{{\Phi _0 }}
{{2\pi }}\frac{d}
{{dt}}\left[ {\tfrac{1}
{2}~\tilde \phi _{1} ~{\kern 1pt} e^{j\theta _{1} } e^{j\omega _1 t} } \right] \hfill \\
   {} &  { = \tfrac{1}
{2}\left[ {\frac{{\Phi _0 {\kern 1pt} \tilde \phi _{1} {\kern 1pt} \omega _1 }}
{{2\pi }}\left( {je^{j\theta _{1} } } \right)} \right]e^{j\omega _1 t} }  
\label{eqn:VoltageOmega1}
\end{align}
By dividing Eq. \ref{eqn:CurrentOmega1} by Eq. \ref{eqn:VoltageOmega1}, we can define a \emph{signal admittance}, $Y_d(\omega_1)$, for this degenerate case.

\begin{align}
\label{eqn:degenerateAdmittanceGeneral}
   {Y_d (\omega _1 )} &  =   {\frac{{i_{1} (t)_+}}
{{v_{1 } (t)_+}}}  \\
 &= {\left( {j{\omega _1}{L_{{\rm{3d,0}}}}} \right)^{ - 1}} + {\left( {j{\omega _1}{L_{{\rm{3d,1}}}}} \right)^{ - 1}}
\label{eqn:degenerateAdmittance}
\end{align}
Above, we have defined the following identities.
\begin{equation}
\boxed{
\begin{gathered}
  {\text{Three-wave degenerate case:}} \\ 
\begin{array}{*{20}{c}}
{{L_{{\rm{3d,0}}}}}& = {{L_J}}\\
\\
{{L_{{\rm{3d,1}}}}}& = { - {L_J}\frac{2}{{\tan \left( F \right)}}\left( {\frac{1}{{\delta f}}} \right){e^{ + j\Delta {\theta _{{\rm{3d}}}}}}}\\
\\
{\Delta {\theta _{\rm{3d}}}}& = {2{\theta _1} - {\theta _3}}
\end{array}\\ 
\end{gathered} 
}
\label{eqn:3waveDegenPumpistor}
\end{equation}

The subscript ``3d" denotes the \emph{three-wave degenerate} case.  We identify the Josephson inductance, $L_J$, from Eq. \ref{eqn:JosephsonInductance} for the unperturbed flux ($\Phi_{\rm{ext}} = \Phi_{\rm{dc}}$) and small phase ($\phi \approx 0$) conditions.  We therefore consider ${L_J} = {\Phi _0}/\left[ {2\pi {I_c}\cos \left( F \right)} \right]$ for the remainder of this work.  From these definitions, Eq. \ref{eqn:degenerateAdmittance} shows that the admittance appears as the parallel combination of the Josephson inductance and a perturbation inductance with an ac-flux dependence (\textit{i.e.}, ``the pumpistor"\cite{Sundqvist:2013}). 

\begin{figure}[h!] 
\centering
  \includegraphics[width=3.5 in]{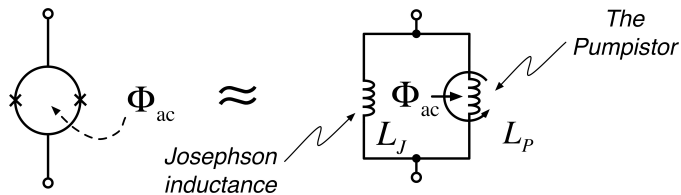} 
   \caption{One may solve for the admittance of a flux-modulated SQUID using series expansions for the super-current. The resulting circuit model appears as the Josephson inductance in parallel to a flux- (and phase-) dependent, inductance-like impedance.}
  \label{Fig2}
 \end{figure}

Note that this extra inductance, $L_{\text{3d,1}}$, has a dependence on the effective pump phase, $\Delta \theta_{\rm{3d}}$. Depending on the value of $\Delta \theta_{\rm{3d}}$,  the inductance $L_{\text{3d,1}}$ has both real and imaginary contributions, which may be either positive or negative.  Our amplifier topology will be able to supply signal gain when $L_{\text{3d,1}}$ has a substantial negative and real impedance.  This depicts the mechanism which allows the SQUID to inject power back into the external circuit at the signal frequency.  A diagram of this equivalent circuit is demonstrated in Fig. \ref{Fig2}.

Here we have treated the degenerate case to first order both in pump flux and in signal phase. We recover the Josephson inductance in combination with a component representing the perturbation to the signal response.  This extra impedance, as defined by its frequency dependence, is an inductor. However, its phase dependence allows it to take on complex amplitudes.  

It is important to point out that, mathematically, this relation only holds at \emph{precisely} the frequency $\omega_1=\omega_3/2$.  When this condition is not met, we need to resort to the general form of the \emph{nondegenerate} case, which we shall treat in sections \ref{sec:generalNondegenerate} and \ref{sec:nondegen3wave}.  

Now, we consider some saturation arguments for this three-wave degenerate case.

\subsection{Saturation of the pump flux for the three-wave degenerate case}

As in the theory of mixers and other nonlinear devices, the nonlinear properties of the driven SQUID lead also to saturation effects.  These effects include the amplitude-dependent modifications of the Josephson inductance, as well as the \emph{gain compression} of the incremental response.  

If we extend the degenerate treatment as in Eq. \ref{eqn:degenerateAdmittanceGeneral}, we can find higher-order parallel inductance terms by expanding the ``flux'' term to higher orders in ac flux.  Taking the series expansion to third-order, we find the following extension to Eq. \ref{eqn:degenerateAdmittance}.

\begin{align}
 \label{eqn:3DegL0Saturating}
{{L_{{\rm{3d}},{\rm{0}}}}}& = {{L_J}}\\
  \label{eqn:3DegL1Saturating}
{{L_{3{\rm{d}},{\rm{1}}}}}& = { - {L_J}\frac{2}{{\tan \left( F \right)}}\left( {\frac{1}{{\delta f}}} \right){e^{ + j\Delta {\theta _{{\rm{3d}}}}}}}\\
   \label{eqn:3DegL2Saturating}
 {{L_{3{\rm{d}},{\rm{2}}}}}& = { - 4{L_J}{{\left( {\frac{1}{{\delta f}}} \right)}^2}}\\
    \label{eqn:3DegL3Saturating}
 {{L_{3{\rm{d}},{\rm{3}}}}}& = {{L_J}\frac{{16}}{{\tan \left( F \right)}}{{\left( {\frac{1}{{\delta f}}} \right)}^3}{e^{ + j\Delta {\theta _{{\rm{3d}}}}}}}\\
  \label{eqn:3DegAngleSaturating}
 {\Delta {\theta _{{\rm{3d}}}}}& = {2{\theta _1} - {\theta _3}}
\end{align}

We find that the parallel inductance terms corresponding to the even powers of ac flux contribute to modifying the standard Josephson inductance.  Meanwhile, the odd powers modify the phase-dependent term.   Knowing that $L_{\text{d,1}} $ is responsible for gain, we can compare it to its higher-order correction, $L_{\text{d,3}}$. So by equating $| L_{\text{d,1}} | $ to $| L_{\text{d,3}} |$ we can estimate the pump ac-flux amplitude ``intercept point."    This is only a rough estimate of saturation, and the effects of gain compression would start to become apparent at ac-fluxes considerably smaller than this.  To ensure that operation is far from this condition, we would say that the following should always be true.
\begin{equation}
\frac{{\Phi _{{\text{ac}}} }}
{{\Phi _{\text{0}} {\kern 1pt} }} \ll \frac{{2 \sqrt{2}}}
{\pi } \approx 0.90
\end{equation} 
This is not a particularly useful constraint, as we already knew that we wish to keep the total external flux below $\Phi_0/2$.  However, we could say that this constraint reinforces the notion that, for properly linearized behavior, $\Phi_{\text{ac}}$ should be maintained at some small fraction of $\Phi_0$. 

\subsection{Saturation in the signal phase (or voltage) for the three-wave degenerate case}
\label{sec:PhaseSaturation}

If we instead now expand the \emph{phase term} of Eq. \ref{SQUIDCurrentItemized} to higher order, we can estimate nonlinear effects due to the magnitude of the \emph{signal phase}.  Here, we assume sin$(\phi) \approx \phi - \frac{1}{6} \phi^3$.  If we again combine the terms which occur at $e^{ j\omega _{1} t }$, we find the 3rd-order correction to the $\Phi_{\text{ac}}$-independent term, $L_{\text{3d,0}}$, to be the following.
\begin{align}
{\raise0.7ex\hbox{$1$} \!\mathord{\left/
 {\vphantom {1 {{L_{{\rm{3d,0}}}}}}}\right.\kern-\nulldelimiterspace}
\!\lower0.7ex\hbox{${{L_{{\rm{3d,0}}}}}$}} \to {\raise0.7ex\hbox{$1$} \!\mathord{\left/
 {\vphantom {1 {{L_J}}}}\right.\kern-\nulldelimiterspace}
\!\lower0.7ex\hbox{${{L_J}}$}}\left( {1 - \frac{{\tilde \phi _{1}^2}}{8}} \right)
\label{eqn:L0With3rdOrderPhase}
\end{align}
We also find a 3rd-order correction to the $L_{\text{3d,1}}$ inductance term, which was the term inversely proportionate to $\delta f$.

\begin{align}
{\raise0.7ex\hbox{$1$} \!\mathord{\left/
 {\vphantom {1 {{L_{{\rm{3d}},{\rm{1}}}}}}}\right.\kern-\nulldelimiterspace}
\!\lower0.7ex\hbox{${{L_{{\rm{3d}},{\rm{1}}}}}$}} \to {\raise0.7ex\hbox{$1$} \!\mathord{\left/
 {\vphantom {1 {{L_{{\rm{3d}},{\rm{1}}}}}}}\right.\kern-\nulldelimiterspace}
\!\lower0.7ex\hbox{${{L_{{\rm{3d}},{\rm{1}}}}}$}}\left[ {1 + \tilde \phi _{1}^2\left( {\frac{1}{{24}}{e^{j2\Delta {\theta _{{\rm{3d}}}}}} - \frac{1}{8}} \right)} \right]
\label{eqn:degenerate_L1_saturation}
\end{align}
To estimate an ``intercept point'' due to saturation of the phase amplitude, we can take the maximum of the corrected $1/L_{\text{3d,1}}$ of Eq. \ref{eqn:degenerate_L1_saturation}, at $\Delta \theta_{\rm{3d}}={\pi}/{2}$.  It is straightforward to see that the contribution of $\tilde \phi_{1}^2$ should be negligible when the following is true.

\begin{equation}
\left| {\tilde {\phi _{1}}} \right| \ll \sqrt 6 
\end{equation}

As in the previous consideration of the nonlinearity due to $\Phi_{\text{ac}}/\Phi_0$, this is not particularly a remarkable constraint.  The phase amplitude $\sqrt{6}$ is obviously already a large fraction of $\pi$.  It only reinforces the point that $|\tilde \phi_1|$ should be quite small compared to this value.  Perhaps, though, it is worthwhile to point out this limit also corresponds directly to a limit on the junction \emph{voltage}, by way of the ac Josephson effect.

\begin{equation}
\left| {{V_{1}}} \right| = {\tilde \phi _{1}}\left( {\frac{{{\Phi _{\rm{0}}}{\omega _1}}}{{2\pi }}} \right) \ll \frac{{\sqrt 2 }}{\pi }{\Phi _{\rm{0}}}{\omega _1}
\end{equation}

\begin{figure}[h!] 
  \centering
   \includegraphics[width=4.25 in]{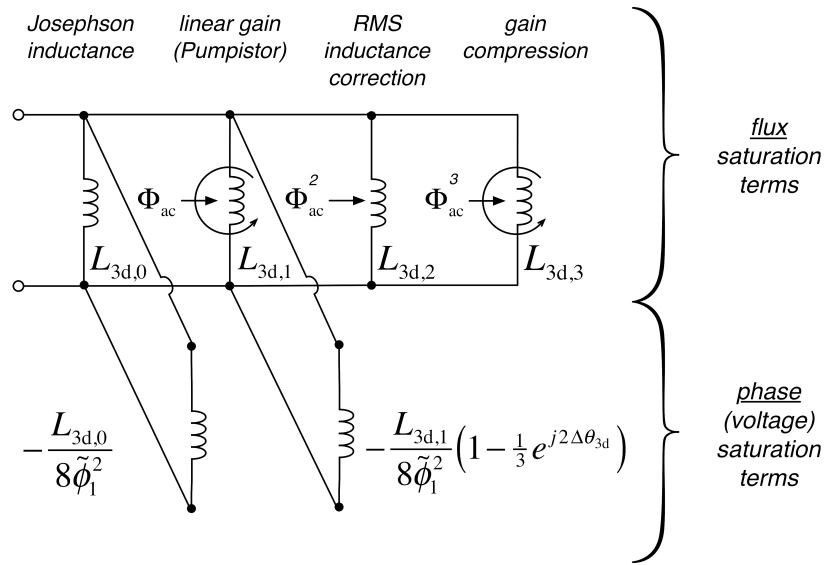} 
   \caption{The admittance expansion to higher order, both in external flux and in junction phase, gives a more complete model for the three-wave degenerate case.  The even harmonics in the flux expansion serve to only modify the net inductance value.  The odd harmonics modify the potential gain and phase-sensitivity.  This allows for estimation of the saturation effects.   As it is in the theory of mixers, we see in the ac-flux expansion that the third-order term compresses the gain-providing first-order term.}
   \label{Fig3}
\end{figure}

Concluding discussion of saturation effects due to flux and to signal phase, we turn to Fig.  \ref{Fig3}.  Here, we combine the effects of gain compression into a common model.  As in the theory of mixers, we see that the odd terms in the expansion account for both gain and its saturation.

\clearpage
  
\section{The four-wave degenerate case}
\label{sec:FourWaveDegenerate}

Next, we take interest in the SQUID with zero dc flux.  When the dc flux is zero, the first derivative of inductance as a function of flux is also zero.  We notice from Eq. \ref{eqn:3DegL1Saturating} that $L_{3d,1}$ becomes infinite (an ``open") and no longer contributes to the circuit.   In fact, all of the odd powers of $\Phi_{\text{ac}}$ will disappear from the ``flux'' term of Eq. \ref{SQUIDCurrentItemized}.   The reason for this can be attributed to the symmetric behavior of the unbiased device.  Yet it is still possible to achieve parametric amplification among the even harmonics of the admittance expansion in flux, in a degenerate case without an idler tone distinct from a signal ($\omega_1 = \omega_2$).  In this case one must use \emph{four-wave degenerate} mixing, where \emph{two} pump photons interact with the signal and idler photons.  This condition requires the following.
\bigskip
\begin{center}
{\bf{The four-wave degenerate case:}}
\end{center}
\begin{equation}
\omega _1  + \omega _2  = 2\omega _3 
\end{equation} 

As in the three-wave degenerate case, both idler and signal tones occur at identically the same frequency and we consider only the disturbance of their combined response.  We call this the \emph{signal} ($\omega_1$) response.

When the external magnetic flux is comprised of solely the ac component, we mentioned that the device behaves symmetrically around zero flux.  To find the relevant dynamical response, we need to expand the ``flux'' term of Eq. \ref{SQUIDCurrentItemized} to \emph{2nd-order} for this four-wave case. 
\begin{equation}
\begin{array}{*{20}{c}}
{{I_c}\cos \left( {\pi \;\delta \Phi /{\Phi _0}} \right)}& \approx &{{I_c} - {I_c}\frac{{{\pi ^2}}}{{2\Phi _0^2}}{{\left( {\delta \Phi } \right)}^2}}&{}\\
{}& = &{{I_c} - \frac{{{I_c}}}{2}{{\left( {\delta f} \right)}^2}{{\left[ {\cos ({\omega _3}t + {\theta _3})} \right]}^2}}&{}
\end{array}
\end{equation}

As before, we find the total current at the signal frequency by multiplying our ``flux'' approximation by the phase term approximation.  We use the approximation for the signal phase as in Eq. \ref{eqn:PhaseExpand}.  The resulting signal current, analogous to Eq. \ref{eqn:CurrentOmega1} but with $\omega_1 = \omega_3$, becomes

\begin{equation}
{i_{{1}}}(t)_{+} = \frac{1}{2}{I_c}{\tilde \phi _{1}}\left[ {1 - \frac{1}{4}{{\left( {\delta f} \right)}^2}} \right]{e^{j{\theta _1}}}{e^{j{\omega _1}t}} - \frac{1}{{16}}{I_c}\tilde \phi _{1}^*{\left( {\delta f} \right)^2}{e^{j2{\theta _3} - j{\theta _1}}}{e^{j{\omega _1}t}}
\label{eqn:CurrentOmega1FourWave}
\end{equation}

Considering the small-signal voltage of Eq. \ref{eqn:VoltageOmega1}, we find the signal admittance in the four-wave degenerate case to be
\begin{align}
{{Y_{4{\rm{d}}}}({\omega _1})}& = {\frac{{2{I_c}\pi }}{{j{\omega _1}{\Phi _0}}} - \frac{{{I_c}\pi }}{{j2{\omega _1}{\Phi _0}}}{{\left( {\delta f} \right)}^2} - \frac{{{I_c}\pi }}{{j4{\omega _1}{\Phi _0}}}{{\left( {\delta f} \right)}^2}{e^{j2{\theta _3} - j2{\theta _1}}}}\\
\notag \\
{}& = {{{\left( {j{\omega _1}{L_{4{\rm{d}},{\rm{0}}}}} \right)}^{ - 1}} + {{\left( {j{\omega _1}{L_{{\rm{4d,2a}}}}} \right)}^{ - 1}} + \left( {j{\omega _1}{L_{{\rm{4d,2b}}}}} \right){.^{ - 1}}}
\label{eqn:Admittance4Wave}
\end{align}

In this case, we define the following.

\begin{equation}
\boxed{
\begin{gathered}
  {\text{Four-wave degenerate case:}} \\ 
\begin{array}{*{20}{c}}
{{L_{{\rm{4d,0}}}}}& = &{{L_J}}\\
{}&{}&{}\\
{{L_{{\rm{4d,2a}}}}}& = &{ - 4{L_J}{{\left( {\frac{1}{{\delta f}}} \right)}^2}}\\
{}&{}&{}\\
{{L_{{\rm{4d,2b}}}}}& = &{ - 8{L_J}{{\left( {\frac{1}{{\delta f}}} \right)}^2}{e^{j\Delta {\theta _{{\rm{4d}}}}}}}\\
{}&{}&{}\\
{\Delta {\theta _{{\rm{4d}}}}}& = &{2\left( {{\theta _1} - {\theta _3}} \right)}\\
{}&{}&{}
\end{array}\\ 
\end{gathered} 
}
\end{equation}

So we find that the admittance which is proportionate to $(\delta f)^2$ has both phase-insensitive and phase-sensitive terms.  Note also the dependence on the pump phase in ${\Delta \theta _{{\text{4d}}} } $ is different by $2$ compared to the phase angle ${\Delta \theta _{{\text{3d}}} } $ of Eq. \ref{eqn:3waveDegenPumpistor}.  Also in this four-wave degenerate case, we can produce a negative resistance, and consequently gain, from the $L_{\rm{4d,2b}}$ term by adjusting ${\Delta \theta _{{\text{4d}}} } $ accordingly.

In the following sections, we turn to the more general case of a \emph{nondegenerate} operation.  There, the idler response must now be considered separately from the signal response.

\section{General conditions for nondegenerate parametric effects using the small-signal admittance matrix}
\label{sec:generalNondegenerate}

We now turn to specifically \emph{nondegenerate} mixing conditions.  Here, ``nondegenerate" asserts its standard meaning that all frequency terms under consideration are unique, i.e., $\omega_i  \ne \omega_j $ for all $ j \ne i$.   Where any of our six considered mixing frequencies (Table \ref{tab:freq}) may contribute to a flux-pumped SQUID circuit, we work within our typical small-signal limit using a linearized system of equations.  From this, we will develop an equivalent admittance matrix.

Consider a general voltage response at any of the six frequencies.
\begin{equation}
{v_n}\left( t \right) = {\textstyle{1 \over 2}} ~ {V_{n}}~{e^{j{\omega _n}t}} + {\textstyle{1 \over 2}} ~{V^*_{n}}~{e^{ - j{\omega _n}t}}\quad \quad \quad n \in \{ 1,2...6\} 
\label{eqn:voltForNondegen}
\end{equation}
The amplitude ${v_{0,n}}$ is complex and eliminates the need to introduce a phase angle as we did before in Eq. \ref{eqn:voltageForm}.  Here, Eq. \ref{eqn:voltForNondegen} also demonstrates that we have adopted the electrical engineering convention for complex number, $j$, rather than the physics convention, $i$, since the convention for the phase factors are opposite compared to what one would find in the quantum optics literature.  

We assume ideal, sinusoidal tones.  In this case, the ac Josephson effect gives junction phases from Eq. \ref{eqn:voltForNondegen} as 

\begin{align}
{{\phi _n}(t)}&{ = \frac{{2\pi }}{{{\Phi _0}}}\int {{v_n}(t)dt =  - j\frac{\pi }{{{\Phi _0}{\omega _n}}}~{V_{n}}~{e^{j{\omega _n}t}} + j\frac{\pi }{{{\Phi _0}{\omega _n}}}~{V_n^*}~{e^{ - j{\omega _n}t}}} }
\label{eqn:phiForNondegen}
\end{align}

Meanwhile, currents will also flow at any of the frequencies.
\begin{equation}
{i_n}(t) = {\textstyle{1 \over 2}} ~ {I_{n}}~{e^{j{\omega _n}t}} +  {\textstyle{1 \over 2}} ~ I^*_{n}~ {e^{ - j{\omega _n}t}}\quad \quad \quad n \in \{ 1,2...6\} 
\label{eqn:currentsForNondegen}
\end{equation}
As before, the SQUID current is directly related to the junction phase by the dc Josephson effect as in Eq. \ref{SQUIDCurrentItemized}, while an external flux is again driven at $\omega_3$ as in Eq. \ref{eqn:FluxPerturbation}.  We continue to work in the limit of small junction phase, $\text{sin}[\phi(t)] \approx \phi(t)$.  Furthermore, we know the total junction phase due to all six considered frequencies is the superposition $\phi (t) = \sum\limits_{n = 1}^6 {{\phi _n}(t)} $, with $\phi _n(t)$ taken from Eq. \ref{eqn:phiForNondegen}.  This gives the total SQUID current approximated as the following.
\begin{align}
{i\left( t \right)}&  \approx {\underbrace {\left\{ {{I_c}\,\cos \left[ {F + \delta f\cos \left( {{\omega _3}t + {\theta _{3}}} \right)} \right]} \right\}}_{\text{``flux'' term}}\underbrace {\left( {\sum\limits_{n = 1}^6 {{\phi _n}(t)} } \right)}_{\text{``phase'' term}}} 
\label{eqn:SQUIDCurrentGeneralNonDegen}
\end{align}

We wish to find the contributions of the current at different frequencies, given by the form $i(t) = \sum\limits_{n = 1}^6 {{i_n}(t)}$ as in Eq. \ref{eqn:currentsForNondegen}.   For a given frequency of junction phase, we find the current amplitudes at all frequencies.  To do this we expand the ``flux'' term of Eq. \ref{eqn:SQUIDCurrentGeneralNonDegen} to second order, which provides nontrival mixed current amplitudes at all frequencies.  We next translate junction phase amplitudes into voltage amplitudes by way of Eq. \ref{eqn:phiForNondegen}.   We take advantage of conjugate symmetries to arrive at a simplified, small-signal admittance matrix. Rather than a basis set of physical ports as in a multi-terminal device, here the admittance matrix ``ports'' (indices) represent the frequencies from Table 1.

\begin{equation}
\left( {\begin{array}{*{20}{c}}
{{I_{1}}}\\
{}\\
{I_{2}^*}\\
{}\\
{{I_{4}}}\\
{}\\
{I_{5}^*}\\
{}\\
{{I_{6}}}
\end{array}} \right) = \frac{1}{{j{L_J}}}\left( {\begin{array}{*{20}{c}}
{\frac{{{\epsilon _0}}}{{{\omega _1}}}}&{ - \frac{{\epsilon _1^*}}{{{\omega _2}}}}&{\frac{{{\epsilon _1}}}{{{\omega _4}}}}&{ - \frac{{\epsilon _2^*}}{{{\omega _5}}}}&{\frac{{{\epsilon _2}}}{{{\omega _6}}}}\\
{}&{}&{}&{}&{}\\
{\frac{{{\epsilon _1}}}{{{\omega _1}}}}&{ - \frac{{{\epsilon _0}}}{{{\omega _2}}}}&{\frac{{{\epsilon _2}}}{{{\omega _4}}}}&{ - \frac{{\epsilon _1^*}}{{{\omega _1}}}}&0\\
{}&{}&{}&{}&{}\\
{\frac{{\epsilon _1^*}}{{{\omega _1}}}}&{ - \frac{{\epsilon _2^*}}{{{\omega _2}}}}&{\frac{{{\epsilon _0}}}{{{\omega _4}}}}&0&{\frac{{{\epsilon _1}}}{{{\omega _6}}}}\\
{}&{}&{}&{}&{}\\
{\frac{{{\epsilon _2}}}{{{\omega _1}}}}&{ - \frac{{{\epsilon _1}}}{{{\omega _2}}}}&0&{ - \frac{{{\epsilon _0}}}{{{\omega _5}}}}&0\\
{}&{}&{}&{}&{}\\
{\frac{{\epsilon _2^*}}{{{\omega _1}}}}&0&{\frac{{\epsilon _1^*}}{{{\omega _4}}}}&0&{\frac{{{\epsilon _0}}}{{{\omega _6}}}}
\end{array}} \right)\left( {\begin{array}{*{20}{c}}
{{V_{1}}}\\
{}\\
{V_{2}^*}\\
{}\\
{{V_{4}}}\\
{}\\
{V_{5}^*}\\
{}\\
{{V_{6}}}
\end{array}} \right)
\label{eqn:generalMatrix}
\end{equation}

We do not list in this matrix the pump current amplitude, $I_3$, as it does not include any off-diagonal terms to couple to other frequencies.    

This admittance matrix holds true as long as the pump frequency is larger than the signal frequency ($\omega_3 > \omega_1$) so that the ``three-wave difference idler'' frequency remains positive ($\omega_2>0$).  In the case of $\omega_3 < \omega_1$, some matrix elements appear instead with conjugate quantities.  Similar redefinitions are also necessary if frequency $\omega_5=2 \omega_3 - \omega_1$ were also to become negative.  We consider the conditions ($\omega_2 > 0 $) and ($\omega_5 > 0$) to be the standard situation.

We find again the quiescent Josephson inductance, ${L_J}={\textstyle{{{\Phi _0}} \over {2\pi \,{I_c}\cos \left( F \right)}}}$.  Some new, flux-dependent terms $\epsilon_0$, $\epsilon_1$, and $\epsilon_2$ also appear, which are not indexed by frequency. Rather, their indices indicate the order of the series expansion in flux for which they first become nontrivial.  Their expressions are the following.
\begin{align}
\label{eqn:epsilon0}
\epsilon _0 &= 1 - \frac{1}{4}\delta {f^2}\\
\notag\\
\label{eqn:epsilon1}
\epsilon _1  &= {\frac{{\delta f}}{2}{\rm{tan}}(F){e^{ - j{\theta _3}}}} \\
\notag\\
{{\epsilon_2}}&={\frac{{\delta {f^2}}}{8}{e^{ - j2{\theta _3}}}}
\label{eqn:epsilon2}
\end{align} 

To note, for vanishing $\delta f = \frac{\pi \Phi_{\rm{ac}}}{\Phi_0}$, the limit of $\epsilon_0$ is unity, while $\epsilon_1$ and $\epsilon_2$ tend to zero.

The importance of the matrix equation (Eq. \ref{eqn:generalMatrix}) should be emphasized.  This tells us the response of a flux-pumped SQUID between all relevant frequency components, but yet it can be used in the same form as any other n-port admittance matrix from circuit theory.  So for this very general degenerate case, we may now consider a large number of three-wave and four-wave mixing devices, both as negative-resistance amplifiers and as frequency converters.  It further allows us to describe a number of next-order effects which also occur in these devices.  

The elements of the admittance matrix (Eq. \ref{eqn:generalMatrix}), are specifically the \emph{short-circuit} admittance parameters\cite{Chen:1980}.  This is defined as the following,
\begin{equation}
{Y_{kl}} = {\left. {\frac{{{I_k}}}{{{V_l}}}} \right|_{{V_m} = 0,\,m \ne l}}
\end{equation}
where $V_m = 0$ with $m \ne l$  is a condition met by shorting all ports, $m$ other than the port of interest, $l$. 

In the next section, we begin by considering a special case of Eq. \ref{eqn:generalMatrix} where the desired harmonics form a subset of the admittance matrix.  The unwanted components are assumed to be zero (i.e., shorted).  We will then find necessary corrections for when unwanted harmonics are instead \emph{open-circuited}.

\section{The three-wave nondegenerate negative-resistance parametric amplifier}
\label{sec:nondegen3wave}

When the signal frequency under consideration is NOT specifically $\omega_3/2$ or $\omega_3$, the ``degenerate'' conditions of sections \ref{sec:3waveDegen} and \ref{sec:FourWaveDegenerate} break down.  We now return to considerations of three-wave mixing, but for the \emph{nondegenerate} case where $\omega _1  \ne \omega _3 /2$. In this case, it is necessary to provide for the presence of an \emph{idler} junction phase (at $\omega_2$).  The idler comes about due to the nonlinear frequency coupling between the signal and pump terms. A response at the idler frequency need not be induced at the input, nor measured as an output variable, to play an important role as an internal state variable.  

In this section, we consider the following conditions on the signal and idler frequencies.

\begin{center}
{\bf{The three-wave nondegenerate case:}}
\end{center}
\begin{align}
 \omega_2 & =  (\omega_3 - \omega_1)  \ne  \omega_1 
\label{eqn:DegenerateFreq}
\end{align}

We consider the matrix subset of Eq. \ref{eqn:generalMatrix} corresponding to a signal at $\omega_1$ and the idler at $\omega_2$.  The circuit at all other harmonics is assumed to be shorted.

\begin{align}
\left( {\begin{array}{*{20}{c}}
{{I_1}}\\
{I_2^*}
\end{array}} \right) = \frac{1}{{j{L_J}}}\left( {\begin{array}{*{20}{c}}
{\frac{{{\epsilon _0}}}{{{\omega _1}}}}&{ - \frac{{\epsilon _1^*}}{{{\omega _2}}}}\\
{\frac{{{\epsilon _1}}}{{{\omega _1}}}}&{ - \frac{{{\epsilon _0}}}{{{\omega _2}}}}
\end{array}} \right)\left( {\begin{array}{*{20}{c}}
{{V_1}}\\
{V_2^*}
\end{array}} \right)
\label{eqn:paramp3WaveMatrix}
\end{align}
This provides the current and voltage relations directly across the SQUID at the signal and idler frequencies.  Next, we generalize the circuit such that we take into account the possible effects of other generator and load admittances. 

\subsection{Understanding this three-wave nondegenerate model as a circuit topology}
\label{sec:3waveTopology}

\begin{figure}[h!] 
\centering
   \includegraphics[width=4.0 in]{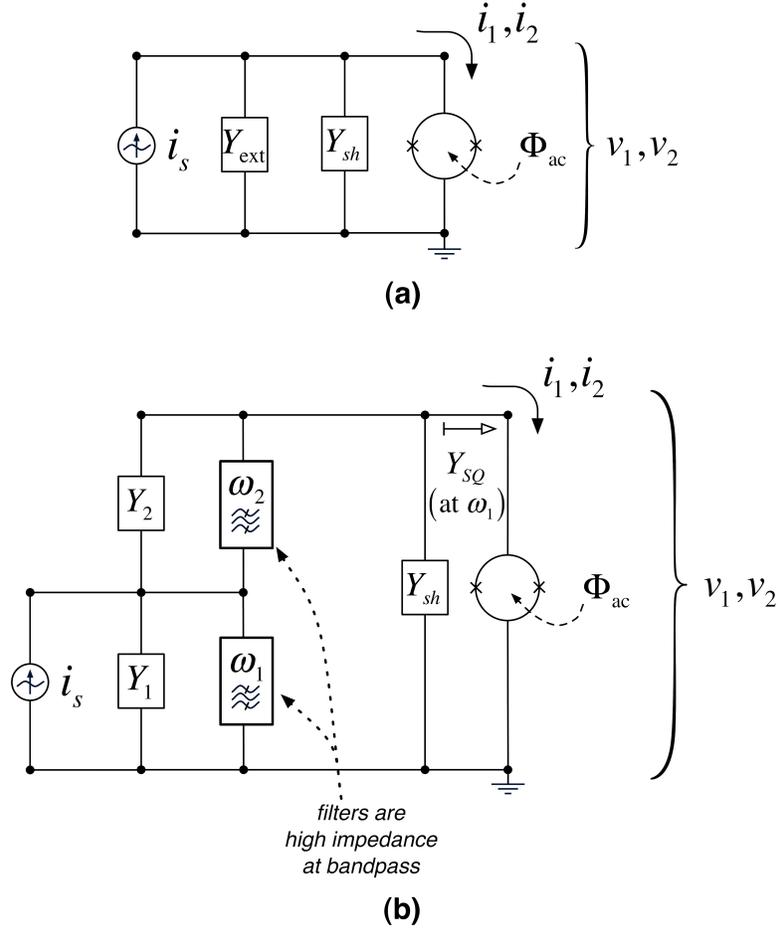} 
   \caption{ { \bf{(a)}} This figure depicts the physical, general circuit considered in this section, containing a flux-pumped SQUID. This circuit accounts for an external source current, $i_s$, as well as external ($Y_{\rm{ext}}$) and local shunt ($Y_{\rm{sh}}$) admittances.  { \bf{(b)}}  It is possible to represent the general circuit in an equivalent way that separates the external loading effects of the circuit at the ``signal'' ($\omega_1$) and ``idler''  ($\omega_2$) frequencies.  This is done by introducing hypothetical, ideal bandpass filters at $\omega_1$ and $\omega_2$.  These filters act open-circuited at their respective frequency, but short-circuited for all other frequencies.  In this representation, the external admittance, $Y_{\rm{ext}}$, is represented at different frequencies by $Y_1$ and $Y_2$. The input admittance ($Y_{\rm{SQ}}$) is then based on the signal voltage response to current $i_1$, which depends upon the idler in a way that may allow for gain. }
   \label{Fig4}
\end{figure}

To conceptualize the system we have just described, consider the flux-pumped SQUID as the primary element of a multi-frequency circuit.  This is depicted in Fig. \ref{Fig4}(a).   We assume this circuit to be sourced by a signal current $i_{s}(t)$ of the form of Eq. \ref{eqn:currentsForNondegen}.  This external current source at $\omega_1$, may be loaded by an external admittance, $Y_{\rm{ext}}$.  The currents $i_{1}(t)$ and $i_{2}(t)$ continue to indicate the currents directly into the SQUID at frequencies $\omega_1$ and $\omega_2$, respectively. We account for either parasitic or intentional admittances directly across the SQUID by the term $Y_{\rm{sh}}$.  We make a distinction between $Y_{\rm{ext}}$ and $Y_{\rm{sh}}$ since the definition of available power from the external source involves only $\text{Re}[Y_{\rm{ext}}]$.  

In Fig.\ref{Fig4}(b), we depict how we can think of the effects of the external load at different frequencies by recasting this circuit in an equivalent representation.  In this case, we separate $Y_{\rm{ext}}$ into distinct impedances $Y_1$ at frequency $\omega_1$ and $Y_2$ at frequency $\omega_2$.  We introduce hypothetical bandpass filters which isolate $Y_1$ and $Y_2$ to their respective frequencies outside of the pumped SQUID.  These ideal filters work by providing a high-impedance (open) at their intended frequencies, while at all other frequencies they serve as a perfect short. This topology ensures that all unwanted frequencies short the SQUID, preventing any voltage at those frequencies to accumulate.  Thus, we are able to reduce the general admittance matrix of Eq. \ref{eqn:generalMatrix} to the much simpler matrix of Eq. \ref{eqn:paramp3WaveMatrix}. While we do not actively source the idler current, we will find that the external admittance at the idler frequency, $Y_2$, effects response of the SQUID at the signal frequency in an important way.

\subsection{The voltage and current ratios of the three-wave nondegenerate parametric amplifier}

We are not quite ready to understand how gain appears in this system.  This nondegenerate case is complicated by the appearance of an idler response distinct from the signal.  For instance, the idler-to-signal voltage ratio $\frac{{V_2^*}}{{V_1}}$ will become important.  To find a relation for ${V_2^*}$, we examine the second line of the matrix equation \ref{eqn:paramp3WaveMatrix}.  While it is clear that we need to solve for ${V_2^*}$, what is ${I_2^*}$?  Unlike the signal response, we are not sourcing or measuring an idler current. The idler current is the result of voltage disturbances at the idler frequency, coupled to the external circuit of the surrounding electrical system.  So we need to specify how the circuit is loaded at the idler frequency, which was why it was necessary to specify some external (conjugate) admittance $Y^*_2$ in the previous section.  In what follows, we complete an analysis of our generalized circuit to solve for the idler voltage and current in terms of the signal.

Regarding the general circuit as depicted in Fig. \ref{Fig4}, we use Kirchhoff's node equations for both the signal and idler. 
\begin{align}
{I_{{s}}} - {V_1} {Y_1}^\prime - {I_1} &= 0\\
 - V_2^* {Y_2}^{\prime *}  - I_2^* &= 0
\end{align}
Above, we defined the grouped admittances ${Y_1}^\prime  = {Y_1} + {Y_{{\rm{sh}}}}$ and ${Y_2}^\prime  = {Y_2} + {Y_{{\rm{sh}}}}$.  To go further, the coupled subsystem of Eq. \ref{eqn:paramp3WaveMatrix} allows us to eliminate $I_1$ and $I^*_2$, giving the following.
\begin{align}
\label{eqn:sigCurrentReducedWithGen}
{{I_s}}&{ = {V_1}\left( {{Y_1}^\prime  + \frac{\epsilon_0}{{j{\omega _1}{L_J}}}} \right) - \frac{{V_2^*\epsilon _1^*}}{{j{\omega _2}{L_J}}}}\\
0&{ = \frac{{{V_1{\epsilon _1}}}}{{j{\omega _1}{L_J}}} + V_2^*\left( {{Y_2}^{\prime *} - \frac{{\epsilon _0}}{{j{\omega _2}{L_J}}}} \right)}
\label{eqn:idlerCurrentReducedWithNoGen}
\end{align}
Eq. \ref{eqn:sigCurrentReducedWithGen} and \ref{eqn:idlerCurrentReducedWithNoGen} now represent the current and voltage response of the generalized circuit depicted in Fig. \ref{Fig4}.  Since the \emph{signal current} is sourced in this model (with amplitude $I_{s}$), what remains to be solved are the voltage disturbances $V_1$ and $V^*_2$.  We define the impedances ${Z_{L1}} = j{\omega _1}{L_J}/{\epsilon _0}$ and ${Z_{L2}} = j{\omega _2}{L_J}/{\epsilon _0}$. The voltage amplitudes are then found to be
\begin{align}
\label{eqn:V1ReducedForCircuit}
{V_1} & = L_J^2{Z_{L1}}{\omega _1}{\omega _2}\left( {{Y_2}^{\prime *}{Z_{L2}} - 1} \right)\left( {\frac{{{I_s}}}{\Delta }} \right) \\
V_2^* &= j{L_J}{Z_{L1}}{Z_{L2}}{\epsilon _1}{\omega _2}\left( {\frac{{{I_s}}}{\Delta }} \right),
\label{eqn:V2StarReducedForCircuit}
\end{align}
where the denominator term, $\Delta$, is proportionate to the determinant formed by the matrix of Eqs. \ref{eqn:sigCurrentReducedWithGen} and \ref{eqn:idlerCurrentReducedWithNoGen}.
\begin{align}
\Delta  = L_J^2{\omega _1}{\omega _2}\left( {{Y_1}{Z_{L1}} + 1} \right)\left( {{Y_2}^{\prime *}{Z_{L2}} - 1} \right) - {Z_{L1}}{Z_{L2}}{\left| {{\epsilon _1}} \right|^2}
\end{align}

When we consider the \emph{voltage ratio} between the idler and signal, the cumbersome denominator cancels, providing the more simple relation 
\begin{align}
\frac{{V_2^*}}{{{V_1}}} = \frac{{{\omega _2}{\epsilon _1}}}{{{\omega _1}{\epsilon _0}}}\frac{1}{{1 + Z_{L2}^*{Y_2}^{\prime *}}}.
\label{eqn:voltageRatio}
\end{align}
Where $ Z_{L2}^*{Y_2}^{\prime *} \ll 1$, we see Eq. \ref{eqn:voltageRatio} go to the limit
\begin{align}
\mathop {\lim }\limits_{Z_{L2}^*{Y_2}^{\prime *} \to 0} \left( {\frac{{V_2^*}}{{{V_1}}}} \right) = \frac{{{\omega _2}{\epsilon _1}}}{{{\omega _1}{\epsilon _0}}} = \frac{{{\omega _2}}}{{2{\omega _1}}}\frac{{\delta f}}{{1 - {\textstyle{1 \over 4}}\delta {f^2}}}{\rm{tan}}(F){e^{ - j{\theta _3}}}.
\label{eqn:limV2with0}
\end{align}
On the other hand, when this admittance quantity becomes large such that  $Z_{L2}^*{Y_2}^{\prime *}  \gg 1$, we see
\begin{align}
\mathop {\lim }\limits_{Z_{L2}^*{Y_2}^{\prime *} \to \infty } \left( {\frac{{V_2^*}}{{{V_1}}}} \right) = 0.
\label{eqn:limV2withInf}
\end{align}
So the voltage of the idler response is of course a function of how well the external circuit is being kept ``open" at the idler frequency, $\omega_2$.

We can also find the idler-to-signal \emph{current ratio}. For this we revisit the system represented by Eq. \ref{eqn:paramp3WaveMatrix}, and divide its second equation by its first.  We substitute the signal and idler voltage amplitudes found in Eq. \ref{eqn:V1ReducedForCircuit} and \ref{eqn:V2StarReducedForCircuit}.   This gives the following.
\begin{align}
\frac{{I_2^*}}{{{I_1}}} = \frac{{{\epsilon _1}}}{{{\epsilon _0}}}\frac{1}{{1 + \frac{1}{{Z_{L2}^*{Y_2}^{\prime *}}} + \frac{{{{\left| {{\epsilon _1}} \right|}^2}}}{{\epsilon _0^2}}\frac{{{\omega _1}}}{{{\omega _2}{Z_{L1}}{Y_2}^{\prime *}}}}}
\end{align}
We can look at the admittance limits of the current ratio as well.  When the external admittance is small, we see
\begin{align}
\mathop {\lim }\limits_{Z_{L2}^*{Y_2}^{\prime *} \to 0} \left( {\frac{{I_2^*}}{{{I_1}}}} \right) = 0.
\label{eqn:limI2with0}
\end{align}
Conversely, when external admittance is large, we see
\begin{align}
\mathop {\lim }\limits_{Z_{L2}^*{Y_2}^{\prime *} \to \infty } \left( {\frac{{I_2^*}}{{{I_1}}}} \right) = \frac{{{\epsilon _1}}}{{{\epsilon _0}}} = \frac{1}{2}\frac{{\delta f}}{{1 - {\textstyle{1 \over 4}}\delta {f^2}}}{\rm{tan}}(F){e^{ - j{\theta _3}}}.
\label{eqn:limI2withInf}
\end{align}

These limits are intuitive.  We can see the idler current will be inhibited when the external circuit is comparatively more ``open,'' representing a small external admittance.  Note the similar behavior indicated between Eq. \ref{eqn:limV2with0} and \ref{eqn:limI2with0}, as well as between Eq. \ref{eqn:limV2withInf} and \ref{eqn:limI2withInf}.

These quantities depict the response of the circuit at the idler frequency, $\omega_2$, relative to the circuit behavior at the signal frequency, $\omega_1$.  We will now utilize this understanding in the next section to find how this system acts as a negative-resistance amplifier. 
\subsection{The input impedance of the nondegenerate three-wave parametric amplifier}
\label{sec:shortCircuitedCase}

To understand how this system works as an amplifier, we must find how it provides a negative resistance at the signal frequency.  To this end, we seek to find the input admittance as seen at $\omega_1$.  

The input admittance as seen into the device at the signal frequency we can say is $Y_{SQ} = {{{I_1}} \mathord{\left/
 {\vphantom {{{I_1}} {{V_1}}}} \right. \kern-\nulldelimiterspace} {{V_1}}}$, giving
\begin{align}
{Y_{SQ}} = \frac{{{I_1}}}{{{V_1}}} = \frac{1}{{{Z_{L1}}}} - \frac{{V_2^*}}{{{V_1}}}\frac{{\epsilon _1^*}}{{{\epsilon _0}}}\frac{1}{{{Z_{L1}}}}
\label{eqn:3waveInputAdmit}
\end{align}
Recall that $\epsilon_0 \approx 1$ to first order.  To interpret Eq. \ref{eqn:3waveInputAdmit} as an \emph{impedance}, this is the Josephson inductance again in parallel to some other term.  To find this other term, which is represented (as an admittance) by the second term on the right-hand side of Eq. \ref{eqn:3waveInputAdmit}, we must incorporate the ratio $\frac{{V_2^*}}{{V_1}}$ from Eq. \ref{eqn:voltageRatio}. Substituting this term into Eq. \ref{eqn:3waveInputAdmit}, we arrive at
\begin{align}
{{Y_{SQ}}}&{ = \frac{1}{{{Z_{L1}}}} - \frac{{{{\left| {{\epsilon _1}} \right|}^2}}}{{{\epsilon ^2}_0}}\frac{1}{{{Z_{L1}}}}\frac{1}{{1 + {Z^*}_{L2}{Y_2}^{\prime *}}}}\\
\notag\\
{}&{ = {{\left( {j{\omega _1}{L_{{\rm{n}},{\rm{0}}}}} \right)}^{ - 1}} + {{\left( {j{\omega _1}{L_{{\rm{n}},{\rm{2}}}}} \right)}^{ - 1}}.}
\end{align}
We have therefore represented the input admittance again as inductive terms.  We determined a parallel inductance model before, in the \emph{degenerate} case, but here the dependence on the \emph{pump phase is no longer present}.  This \emph{nondegenerate} amplifier, therefore, is \emph{phase insensitive}.  The following terms for inductances are used.

\begin{equation}
\boxed{
\begin{gathered}
  {\text{Three-wave nondegenerate case:}} \\ 
\begin{array}{*{20}{c}}
{{L_{{\rm{n}},{\rm{0}}}}}& = &{{L_J}/{\varepsilon _0}}\\
{}&{}&{}\\
{{L_{{\rm{n}},{\rm{2}}}}}& = &{ - \frac{{{L_J}}}{{{{\left| {{\varepsilon _1}} \right|}^2}}}\left( {{\varepsilon _0} - j{\omega _2}{L_J}{Y_2}^{\prime *}} \right)}\\
{}&{}&{}
\end{array}
\end{gathered} 
\label{eqn:nondegLn}
}
\end{equation}

\begin{figure}[h!] 
\centering
  \includegraphics[width=3.5 in]{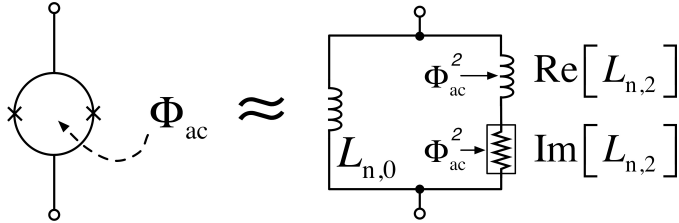} 
   \caption{This figure depicts the equivalent signal impedance of the flux-pumped SQUID in the three-wave \emph{nondegenerate} case.  The constituent inductances are given by Eq. \ref{eqn:nondegLn}.  In the limit that the ac-flux is small such that $\gamma_0 \approx 1$, the inductance $L_{\rm{n,0}}$ is simply the quiescent Josephson inductance and $L_{\rm{n,2}} \propto \Phi _{{\rm{ac}}}^2$.  The $L_{\rm{n,2}}$ acquires an imaginary component due to the external (real) admittance at the idler frequency.  A positive, imaginary \emph{inductance} is a \emph{negative, real impedance}, which may therefore provide gain.  }
   \label{Fig5}
\end{figure}

Above, the ``$L_{\rm{n,0}}$'' inductance is once again simply the Josephson inductance in the small-signal limit.  The ``$L_{\rm{n,2}}$'' inductance, in parallel to $L_{\rm{n,0}}$, contains two terms which are both proportionate to ${{{\left| {{\epsilon _1}} \right|}^{-2}}}$.  The first term is negative and simply modifies the net inductance by a small correction, making the net inductance appear bigger as the ac flux increases.  The second term of $L_{\rm{n,2}}$ depends on ${Y_2}^{\prime *}$ in an important way, providing the possibility for gain in this scenario.  If ${Y_2}^{\prime *}$ has a real and positive component, this allows the impedance represented by $L_{\rm{n,2}}$ to acquire a \emph{negative} and real component.  Therefore the ${Y_2}^{\prime *}$ term of $L_{\rm{n,2}}$ acts as an active impedance converter, allowing the impedance external to the SQUID at the idler frequency to appear, transformed, at the signal frequency.  We may think of the input admittance (or input impedance $Z_{SQ} = {\raise0.7ex\hbox{$1$} \!\mathord{\left/
 {\vphantom {1 {{Y_{{\rm{in,1}}}}}}}\right.\kern-\nulldelimiterspace}
\!\lower0.7ex\hbox{${{Y_{{{SQ}}}}}$}}$) directly into the three-wave nondegenerate pumped SQUID as depicted in Fig. \ref{Fig5}.

We comment on the \emph{frequency dependence} of $L_{\rm{n,2}}$.  If we subscribe to axiomatic circuit theory\cite{Chua:1969, Chua:1987, Chua:2003}, our linearized inductances should have a dependence strictly proportional to $j \omega$.  The second term in $L_{\rm{n,2}}$, which is the same term that may act as a negative resistance, also contains an extra factor, $j \omega_2 = j (  \omega_3 - \omega_1 )$.  This gives a maximum of the product $\omega_1 \omega_2$ at $\omega_3 /2$, which for this reason is why $\omega_3/2$ is the frequency of maximum parametric amplification in a three-wave nondegenerate amplifier.  So between an uncommon frequency dependence and negative-resistance behavior, it may be logical to consider this second term of $L_{\rm{n,2}}$ as relating to something other than an inductance.  We choose keep the terminology of an inductance only for consistency. 

To conclude this section, we repeat that we have found the \emph{negative resistance} that provides gain in this three-wave nondegenerate amplifier.   This appears in the imaginary component of the term  $L_{\rm{n,2}}$ from Eq. \ref{eqn:nondegLn}. Although the frequency mixing between the idler and signal is provided for by the pump, the negative resistance occurs as an effect of mapping the idler's external (real) load admittance back onto the signal as a negative resistance.    

\subsection{The three-wave nondegenerate amplifier: transducer gain}

As we have described a system loaded by specific impedances, we can also specify the \emph{transducer gain} of the device.  The transducer gain is the ratio of the output power to the available input power.  We consider the source admittance as $Y_1$. For the \emph{(rms) available input power} at the signal frequency, we find 
\begin{align}
{P_{{\rm{a,1}}}} &= \frac{{{I_{s}}^2}}{{8~{\mathop{\rm Re}\nolimits} [{Y_1}]}}.
\end{align}
We consider the output signal to be reflected back onto the input admittance, such that we say the \emph{(rms) output power} is
\begin{align}
{P_{{\rm{o,1}}}} &= \frac{{{V_1}^2}}{2~}{\mathop{\rm Re}\nolimits} [{Y_1}].
\end{align}
The transducer gain is then 
\begin{align}
{G_T} = \frac{{{P_{{\rm{o}},{\rm{1}}}}}}{{{P_{{\rm{a}},{\rm{1}}}}}} = \frac{{4{V_1}^2{{\left( {{\rm{Re}}[{Y_1}]} \right)}^2}}}{{{I_s}^2}} = 4\frac{{{{\left( {{\rm{Re}}[{Y_1}]} \right)}^2}}}{{{{\left| {{Y_1}^\prime  + {Y_{SQ}}} \right|}^2}}}.
\end{align}
This can be expressed as
\begin{align} 
{G_T} = \frac{{4\,{\rm{Re}}{{[{Y_1}]}^2}}}{{{{\left( {\frac{1}{{{\omega _1}{L_{{\rm{n}},{\rm{0}}}}}} + \frac{{{\rm{Re}}\left[ {{L_{{\rm{n}},{\rm{2}}}}} \right]}}{{{\omega _1}{{\left| {{L_{{\rm{n}},{\rm{2}}}}} \right|}^2}}} - {\rm{Im}}\left[ {{Y_1}^\prime } \right]} \right)}^2} + {{\left( {\frac{{{\rm{Im}}\left[ {{L_{{\rm{n}},{\rm{2}}}}} \right]}}{{{\omega _1}{{\left| {{L_{{\rm{n}},{\rm{2}}}}} \right|}^2}}} - {\rm{Re}}\left[ {{Y_1}^\prime } \right]} \right)}^2}}}
\end{align}
where ${L_{\rm{n,0}}}$ and ${L_{\rm{n,2}}}$ are from Eq. \ref{eqn:nondegLn}.   
    
\subsection{Adding open-circuited terms}

As an \emph{admittance} model, as opposed to an \emph{impedance} model, the ideal case is for all non-intentional harmonics to be subject to an infinite admittance external to the pumped SQUID (e.g., to have a shorted external load at frequencies other than the signal and idler).  This prevents voltages at these other frequencies from accumulating across the SQUID, thereby removing their influence from the admittance matrix and the resulting mixed currents.  Conversely, when the external impedance is nontrivial at other frequencies, other harmonics will modify the description we have just presented.  

Here, we treat the case opposite from before, where we now consider frequencies other than $\omega_1$ and $\omega_2$ to be \emph{open-circuited}.  Therefore, we consider the last three rows of the admittance matrix of Eq. \ref{eqn:generalMatrix} to represent no current flow, setting currents $I_4$, $I^*_5$, and $I_6$ to zero.  We solve for the voltage amplitudes of these harmonics, which are now nontrivial. We substitute these voltage amplitudes into the expressions for current at the signal ($\omega_1$) and idler ($\omega_2$) frequencies.  To reach a manageable solution, we assume the limiting conditions $\epsilon_0 \approx 1$ and $\epsilon_2 \approx 0$ for currents at $\omega_4$, $\omega_5$, and $\omega_6$.   If we keep terms up to $\delta f^2$, we find the signal-idler subset matrix has the simple form,

\begin{equation}
\left( {\begin{array}{*{20}{c}}
{{I_1}}\\
{I_2^*}
\end{array}} \right) = \frac{1}{{j{L_J}}}\left( {\begin{array}{*{20}{c}}
{\frac{{{\varepsilon _0}}}{{{\omega _1}}}\left[ {1 - \frac{{{{\left| {{\varepsilon _1}} \right|}^2}}}{{{\varepsilon _0}}}} \right]}&{ - \frac{{\varepsilon _1^*}}{{{\omega _2}}}}\\
{\frac{{{\varepsilon _1}}}{{{\omega _1}}}}&{ - \frac{{{\varepsilon _0}}}{{{\omega _2}}}\left[ {1 - \frac{{{{\left| {{\varepsilon _1}} \right|}^2}}}{{{\varepsilon _0}}}} \right]}
\end{array}} \right)\left( {\begin{array}{*{20}{c}}
{{V_1}}\\
{V_2^*}
\end{array}} \right).
\label{eqn:paramp3WaveMatrixOpen}
\end{equation}
We find this system identical to that of Eq. \ref{eqn:paramp3WaveMatrix}, except for the multiplicative correction factor in square brackets, ${\left[ {1 - \frac{{{{\left| {{\varepsilon _1}} \right|}^2}}}{{{\varepsilon _0}}}} \right]}$, appearing in the two matrix elements of the main diagonal.  This correction factor may become significant even for reasonably small $\delta f$ as the dc flux, $F$, approaches $\pi / 2$.  This is the notable difference between this open-circuited case and the short-circuited case we treated in section \ref{sec:shortCircuitedCase}.

\begin{figure}[h!] 
 \includegraphics[width=3.5 in]{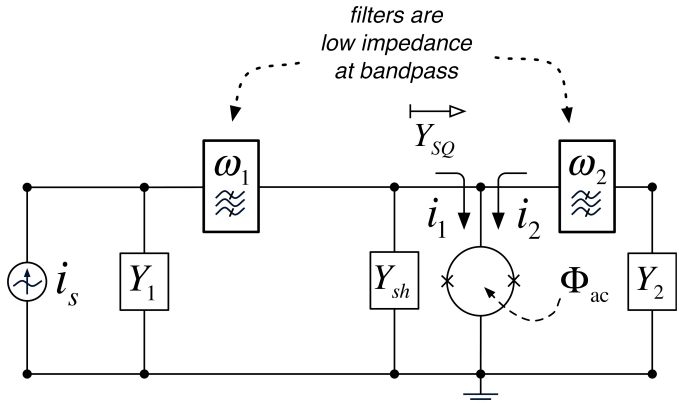} 
 \centering
   \caption{This figure demonstrates the parametric interaction of the {\textbf{open-circuited}} SQUID at the signal and idler frequencies.   As opposed to section \ref{sec:3waveTopology}, these \emph{series} bandpass filters are now \emph{zero} impedance at bandpass and blocking at all other frequencies.  This works in such a way that frequencies other than $\omega_1$ and $\omega_2$ now present an open circuit to the pumped SQUID.  Therefore the voltage across the SQUID is not necessarily zero at these other frequencies.  These additional mixing effects can be mapped onto a \emph{modified} subsystem between signal and idler, which is the 2x2 matrix of Eq. \ref{eqn:paramp3WaveMatrixOpen}.  }
   \label{Fig6}
\end{figure}

We illustrate the open-circuited case as an equivalent circuit in Fig. \ref{Fig6}. We depict signal and idler circuits now directly in parallel to the pumped SQUID.  As opposed to the short-circuited case depicted in Fig. \ref{Fig4}(b), here the ideal filters are accomplished \emph{in series} such that only the permitted frequency is allowed to pass, while all other frequencies see an open-circuit.

In this section, we have determined the response of the three-wave \emph{nondegenerate} amplifier as an impedance model.  This is analogous to the ``pumpistor'' models we found for the three-wave and four-wave degenerate cases treated in sections \ref{sec:3waveDegen} and \ref{sec:FourWaveDegenerate}.  A notable difference in this nondegenerate case is that the external admittance at the idler frequency now determines the negative resistance.  As can be seen by Eq. \ref{eqn:nondegLn}, for a negative resistance to occur at the signal frequency, it is necessary that the circuit external to the SQUID at the idler frequency appear as a positive and real admittance.  By treating both a ``short-circuited" and an ``open-circuited'' model, we found that a finite external admittance at harmonics other than the signal and idler frequencies may also affect amplifier performance.

\section{Conclusions}

In conclusion, we have substantially extended the equivalent impedance models of a flux-pumped SQUID which we first put forth for the three-wave degenerate case\cite{Sundqvist:2013}.  For all the general classes of parametric driving, a flux-pumped SQUID can be described at the signal frequency as a Josephson inductance in parallel to an effective, flux-dependent circuit element, ``the pumpistor.''  Parametric amplification can be intuitively understood within this framework, as the pumpistor impedance manifests in whole or in part as a negative resistance.  These models allow the tools and techniques of circuit theory and radio-frequency engineering to be utilized for these applications.  

We reviewed three-wave degenerate pumping, which explains why gain in this case should be \emph{phase senstive} between the signal and pump.  For this case, we also extended our impedance approximation to demonstrate how the SQUID saturates both by pump flux and by junction phase (or voltage).  We also depicted the four-wave degenerate case which is appropriate when the device is biased with zero-flux.  Here, the pumpistor element is inversely proportionate to the \emph{square} of the ac flux. We found this case also to be phase sensitive, but with a slightly different signal-to-pump difference than in the three-wave degenerate case. 

We also depicted nondegenerate pumping in a very general sense, using a matrix equation formalism. This formalism accounts for the presence of one or up to four ``idler'' frequencies which occur as mixing tones between the pump and the signal response. Many three- and four-wave nondegenerate parametric phenomena can be interpreted from this matrix, including effects such as frequency up- and down-conversion.  Using a subset of these matrix equations, we treated the three-wave nondegenerate amplifier, where the signal and single idler are considered.  By solving for an idler distinct from the signal, we found that the pumpistor impedance was now \emph{phase insensitive}.  We found the negative resistance responsible for gain was now dependent on the external circuit admittance at the idler frequency. With regards to the other, higher harmonics, we treated the three-wave nondegenerate amplifier in both the ``short-circuited'' and ``open-circuited'' approximations. While all of these models operate under a classical, circuit-theoretic framework rather than a quantum optics framework, they should be of great benefit for future designs of experiments using superconducting circuits for quantum information purposes.


\section*{Competing interests}
  The authors declare that they have no competing interests.

\section*{Author's contributions}
  KMS derived most of the equations.  Both authors developed the concept and wrote this paper together.

\section*{Acknowledgements}
  We acknowledge support from the EU through the ERC and the projects SOLID, SCALEQIT, and PROMISCE, as well as from the Swedish Research Council and the Wallenberg Foundation.  We are also grateful for fruitful discussions with Christopher M. Wilson, Seckin Kinta\c{s}, Micha{\"e}l Simoen, Philip Krantz, Martin Sandberg, and Jonas Bylander.


\bibliographystyle{bmc-mathphys} 


\begin{thebibliography}{32}
\ifx \bisbn   \undefined \def \bisbn  #1{ISBN #1}\fi
\ifx \binits  \undefined \def \binits#1{#1}\fi
\ifx \bauthor  \undefined \def \bauthor#1{#1}\fi
\ifx \batitle  \undefined \def \batitle#1{#1}\fi
\ifx \bjtitle  \undefined \def \bjtitle#1{#1}\fi
\ifx \bvolume  \undefined \def \bvolume#1{\textbf{#1}}\fi
\ifx \byear  \undefined \def \byear#1{#1}\fi
\ifx \bissue  \undefined \def \bissue#1{#1}\fi
\ifx \bfpage  \undefined \def \bfpage#1{#1}\fi
\ifx \blpage  \undefined \def \blpage #1{#1}\fi
\ifx \burl  \undefined \def \burl#1{\textsf{#1}}\fi
\ifx \doiurl  \undefined \def \doiurl#1{\textsf{#1}}\fi
\ifx \betal  \undefined \def \betal{\textit{et al.}}\fi
\ifx \binstitute  \undefined \def \binstitute#1{#1}\fi
\ifx \binstitutionaled  \undefined \def \binstitutionaled#1{#1}\fi
\ifx \bctitle  \undefined \def \bctitle#1{#1}\fi
\ifx \beditor  \undefined \def \beditor#1{#1}\fi
\ifx \bpublisher  \undefined \def \bpublisher#1{#1}\fi
\ifx \bbtitle  \undefined \def \bbtitle#1{#1}\fi
\ifx \bedition  \undefined \def \bedition#1{#1}\fi
\ifx \bseriesno  \undefined \def \bseriesno#1{#1}\fi
\ifx \blocation  \undefined \def \blocation#1{#1}\fi
\ifx \bsertitle  \undefined \def \bsertitle#1{#1}\fi
\ifx \bsnm \undefined \def \bsnm#1{#1}\fi
\ifx \bsuffix \undefined \def \bsuffix#1{#1}\fi
\ifx \bparticle \undefined \def \bparticle#1{#1}\fi
\ifx \barticle \undefined \def \barticle#1{#1}\fi
\ifx \bconfdate \undefined \def \bconfdate #1{#1}\fi
\ifx \botherref \undefined \def \botherref #1{#1}\fi
\ifx \url \undefined \def \url#1{\textsf{#1}}\fi
\ifx \bchapter \undefined \def \bchapter#1{#1}\fi
\ifx \bbook \undefined \def \bbook#1{#1}\fi
\ifx \bcomment \undefined \def \bcomment#1{#1}\fi
\ifx \oauthor \undefined \def \oauthor#1{#1}\fi
\ifx \citeauthoryear \undefined \def \citeauthoryear#1{#1}\fi
\ifx \endbibitem  \undefined \def \endbibitem {}\fi
\ifx \bconflocation  \undefined \def \bconflocation#1{#1}\fi
\ifx \arxivurl  \undefined \def \arxivurl#1{\textsf{#1}}\fi
\csname PreBibitemsHook\endcsname

\bibitem{Caves:1982}
\begin{barticle}
\bauthor{\bsnm{Caves}, \binits{C.M.}},
\bjtitle{Phys. Rev. D}
\bvolume{26}(\bissue{8}),
\bfpage{1817}--\blpage{1839}
(\byear{1982})
\end{barticle}
\endbibitem

\bibitem{Feldman:1975}
\begin{barticle}
\bauthor{\bsnm{Feldman}, \binits{M.J.}},
\bauthor{\bsnm{Parrish}, \binits{P.T.}},
\bauthor{\bsnm{Chiao}, \binits{R.Y.}}
\bjtitle{J. Appl. Phys.}
\bvolume{46}(\bissue{9}),
\bfpage{4031}--\blpage{4042}
(\byear{1975})
\end{barticle}
\endbibitem

\bibitem{Taur:1977}
\begin{barticle}
\bauthor{\bsnm{Taur}, \binits{Y.}},
\bauthor{\bsnm{Richards}, \binits{P.L.}},
\bjtitle{J. Appl. Phys.}
\bvolume{48}(\bissue{3}),
\bfpage{1321}--\blpage{1326}
(\byear{1977})
\end{barticle}
\endbibitem

\bibitem{Feldman:1977}
\begin{barticle}
\bauthor{\bsnm{Feldman}, \binits{M.J.}},
\bjtitle{J. Appl. Phys.}
\bvolume{48}(\bissue{3}),
\bfpage{1301}--\blpage{1310}
(\byear{1977})
\end{barticle}
\endbibitem

\bibitem{Wahlsten:1977}
\begin{barticle}
\bauthor{\bsnm{Wahlsten}, \binits{S.}},
\bauthor{\bsnm{Rudner}, \binits{S.}},
\bauthor{\bsnm{Claeson}, \binits{T.}},
\bjtitle{Appl. Phys. Lett.}
\bvolume{30},
\bfpage{298}--\blpage{300}
(\byear{1978})
\end{barticle}
\endbibitem

\bibitem{Wahlsten:1978}
\begin{barticle}
\bauthor{\bsnm{Wahlsten}, \binits{S.}},
\bauthor{\bsnm{Rudner}, \binits{S.}},
\bauthor{\bsnm{Claeson}, \binits{T.}},
\bjtitle{J. Appl. Phys.}
\bvolume{49}(\bissue{7}),
\bfpage{4248}--\blpage{4263}
(\byear{1978})
\end{barticle}
\endbibitem

\bibitem{Yurke:1988}
\begin{barticle}
\bauthor{\bsnm{Yurke}, \binits{B.}},
\bauthor{\bsnm{Kaminsky}, \binits{P.}},
\bauthor{\bsnm{Miller}, \binits{R.}},
\bauthor{\bsnm{Whittaker}, \binits{E.}},
\bauthor{\bsnm{Smith}, \binits{A.}},
\bauthor{\bsnm{Silver}, \binits{A.}},
\bauthor{\bsnm{Simon}, \binits{R.}},
\bjtitle{Phys Rev Lett}
\bvolume{60}(\bissue{9}),
\bfpage{764}
(\byear{1988})
\end{barticle}
\endbibitem

\bibitem{Yurke:1989}
\begin{barticle}
\bauthor{\bsnm{Yurke}, \binits{B.}},
\bauthor{\bsnm{Corruccini}, \binits{L.R.}},
\bauthor{\bsnm{Kaminsky}, \binits{P.G.}},
\bauthor{\bsnm{Rupp}, \binits{L.W.}},
\bauthor{\bsnm{Smith}, \binits{A.D.}},
\bauthor{\bsnm{Silver}, \binits{A.H.}},
\bauthor{\bsnm{Simon}, \binits{R.W.}},
\bauthor{\bsnm{Whittaker}, \binits{E.A.}},
\bjtitle{Phys. Rev. A}
\bvolume{39}(\bissue{5}),
\bfpage{2519}--\blpage{2533}
(\byear{1989})
\end{barticle}
\endbibitem

\bibitem{CastellanosBeltran:2007}
\begin{botherref}
\oauthor{\bsnm{Castellanos-Beltran}, \binits{M.}},
\oauthor{\bsnm{Lehnert}, \binits{K.W.}},
Appl Phys Lett
(2007)
\end{botherref}
\endbibitem

\bibitem{Eichler:2011}
\begin{barticle}
\bauthor{\bsnm{{Eichler, C. and Bozyigit, D. and Lang, C. and Baur, M. and
  Steffen, L. and Fink, J. M. and Filipp, S. and Wallraff, A.}}},
\bjtitle{{Phys. Rev. Lett.}}
\bvolume{107}(\bissue{11}),
\bfpage{113601}
(\byear{2011})
\end{barticle}
\endbibitem

\bibitem{Hatridge:2011}
\begin{barticle}
\bauthor{\bsnm{Hatridge}, \binits{M.}},
\bauthor{\bsnm{Vijay}, \binits{R.}},
\bauthor{\bsnm{Slichter}, \binits{D.H.}},
\bauthor{\bsnm{Clarke}, \binits{J.}},
\bauthor{\bsnm{Siddiqi}, \binits{I.}},
\bjtitle{Phys. Rev. B}
\bvolume{83}(\bissue{13}),
\bfpage{134501}
(\byear{2011})
\end{barticle}
\endbibitem

\bibitem{Yamamoto:2008}
\begin{barticle}
\bauthor{\bsnm{Yamamoto}, \binits{T.}},
\bauthor{\bsnm{Inomata}, \binits{K.}},
\bauthor{\bsnm{Watanabe}, \binits{M.}},
\bauthor{\bsnm{Matsuba}, \binits{K.}},
\bauthor{\bsnm{Miyazaki}, \binits{T.}},
\bauthor{\bsnm{Oliver}, \binits{W.D.}},
\bauthor{\bsnm{Nakamura}, \binits{Y.}},
\bauthor{\bsnm{Tsai}, \binits{J.S.}},
\bjtitle{Appl. Phys. Lett.}
\bvolume{93}(\bissue{4}),
\bfpage{042510}
(\byear{2008})
\end{barticle}
\endbibitem

\bibitem{Wilson:2012}
\begin{bchapter}
\bauthor{\bsnm{Wilson}, \binits{C.M.}},
\bauthor{\bsnm{Duty}, \binits{T.}},
\bauthor{\bsnm{Delsing}, \binits{P.}},
In: \beditor{\bsnm{Dykman}, \binits{M.}} (ed.)
\bbtitle{Fluctuating Nonlinear Oscillators}.
\bpublisher{Oxford University Press},
\blocation{Oxford}
(\byear{2012})
\end{bchapter}
\endbibitem

\bibitem{Sundqvist:2013}
\begin{barticle}
\bauthor{\bsnm{Sundqvist}, \binits{K.M.}},
\bauthor{\bsnm{Kintas}, \binits{S.}},
\bauthor{\bsnm{Simoen}, \binits{M.}},
\bauthor{\bsnm{Krantz}, \binits{P.}},
\bauthor{\bsnm{Sandberg}, \binits{M.}},
\bauthor{\bsnm{Wilson}, \binits{C.M.}},
\bauthor{\bsnm{Delsing}, \binits{P.}},
\bjtitle{Appl. Phys. Lett.}
\bvolume{103}(\bissue{10}),
\bfpage{102603}
(\byear{2013})
\end{barticle}
\endbibitem

\bibitem{Tholen:2007}
\begin{barticle}
\bauthor{\bsnm{Tholen}, \binits{E.A.}},
\bauthor{\bsnm{Ergul}, \binits{A.}},
\bauthor{\bsnm{Doherty}, \binits{E.M.}},
\bauthor{\bsnm{Weber}, \binits{F.M.}},
\bauthor{\bsnm{Gregis}, \binits{F.}},
\bauthor{\bsnm{Haviland}, \binits{D.B.}},
\bjtitle{Appl Phys Lett}
\bvolume{90},
\bfpage{253509}
(\byear{2007})
\end{barticle}
\endbibitem

\bibitem{HoEom:2012}
\begin{barticle}
\bauthor{\bsnm{Ho~Eom}, \binits{B.}},
\bauthor{\bsnm{Day}, \binits{P.K.}},
\bauthor{\bsnm{LeDuc}, \binits{H.G.}},
\bauthor{\bsnm{Zmuidzinas}, \binits{J.}},
\bjtitle{Nat Phys}
\bvolume{8}(\bissue{8}),
\bfpage{623}--\blpage{627}
(\byear{2012})
\end{barticle}
\endbibitem

\bibitem{Mallet:2009}
\begin{barticle}
\bauthor{\bsnm{Mallet}, \binits{F.}},
\bauthor{\bsnm{Ong}, \binits{F.R.}},
\bauthor{\bsnm{Palacios-Laloy}, \binits{A.}},
\bauthor{\bsnm{Nguyen}, \binits{F.}},
\bauthor{\bsnm{Bertet}, \binits{P.}},
\bauthor{\bsnm{Vion}, \binits{D.}},
\bauthor{\bsnm{Esteve}, \binits{D.}},
\bjtitle{Nat Phys}
\bvolume{5}(\bissue{11}),
\bfpage{791}--\blpage{795}
(\byear{2009})
\end{barticle}
\endbibitem

\bibitem{Vijay:2012}
\begin{botherref}
\oauthor{\bsnm{Vijay}, \binits{R.}},
\oauthor{\bsnm{Macklin}, \binits{C.}},
\oauthor{\bsnm{Slichter}, \binits{D.H.}},
\oauthor{\bsnm{Weber}, \binits{S.J.}},
\oauthor{\bsnm{Murch}, \binits{K.W.}},
\oauthor{\bsnm{Naik}, \binits{R.}},
\oauthor{\bsnm{Korotkov}, \binits{A.N.}},
\oauthor{\bsnm{Siddiqi}, \binits{I.}},
Nature
\textbf{490}(77)
(2012)
\end{botherref}
\endbibitem

\bibitem{Flurin:2012}
\begin{barticle}
\bauthor{\bsnm{{Flurin, E. and Roch, N. and Mallet, F. and Devoret, M. H. and
  Huard, B.}}},
\bjtitle{Phys. Rev. Lett.}
\bvolume{109}(\bissue{18}),
\bfpage{183901}
(\byear{2012})
\end{barticle}
\endbibitem

\bibitem{Steffen:2013}
\begin{barticle}
\bauthor{\bsnm{Steffen}, \binits{L.}},
\bauthor{\bsnm{Salathe}, \binits{Y.}},
\bauthor{\bsnm{Oppliger}, \binits{M.}},
\bauthor{\bsnm{Kurpiers}, \binits{P.}},
\bauthor{\bsnm{Baur}, \binits{M.}},
\bauthor{\bsnm{Lang}, \binits{C.}},
\bauthor{\bsnm{Eichler}, \binits{C.}},
\bauthor{\bsnm{Puebla-Hellmann}, \binits{G.}},
\bauthor{\bsnm{Fedorov}, \binits{A.}},
\bauthor{\bsnm{Wallraff}, \binits{A.}},
\bjtitle{Nature}
\bvolume{500}(\bissue{7462}),
\bfpage{319}--\blpage{322}
(\byear{2013})
\end{barticle}
\endbibitem

\bibitem{Abdo:2011}
\begin{barticle}
\bauthor{\bsnm{Abdo}, \binits{B.}},
\bauthor{\bsnm{Schackert}, \binits{F.}},
\bauthor{\bsnm{Hatridge}, \binits{M.}},
\bauthor{\bsnm{Rigetti}, \binits{C.}},
\bauthor{\bsnm{Devoret}, \binits{M.H.}},
\bjtitle{Appl. Phys. Lett.}
\bvolume{99}(\bissue{16}),
\bfpage{162506}
(\byear{2011})
\end{barticle}
\endbibitem

\bibitem{Blackwell:1961}
\begin{bbook}
\bauthor{\bsnm{Blackwell}, \binits{L.A.}},
\bauthor{\bsnm{Kotzebue}, \binits{K.L.}}:
\bbtitle{Semiconductor-Diode Parametric Amplifiers}.
\bpublisher{Prentice Hall},
\blocation{Englewood Cliffs}
(\byear{1961})
\end{bbook}
\endbibitem

\bibitem{Decroly:1973}
\begin{bbook}
\bauthor{\bsnm{Decroly}, \binits{J.C.}}:
\bbtitle{Parametric Amplifiers}.
\bpublisher{Wiley},
\blocation{New York}
(\byear{1973})
\end{bbook}
\endbibitem

\bibitem{Howson:1970}
\begin{bbook}
\bauthor{\bsnm{Howson}, \binits{D.P.}},
\bauthor{\bsnm{Smith}, \binits{R.B.}}:
\bbtitle{Parametric Amplifiers}.
\bpublisher{McGraw-{H}ill},
\blocation{London}
(\byear{1970})
\end{bbook}
\endbibitem

\bibitem{Josephson:1962}
\begin{barticle}
\bauthor{\bsnm{Josephson}, \binits{B.D.}},
\bjtitle{Phys. Lett.}
\bvolume{1},
\bfpage{251}--\blpage{253}
(\byear{1962})
\end{barticle}
\endbibitem

\bibitem{Zagoskin:2011}
\begin{bbook}
\bauthor{\bsnm{Zagoskin}, \binits{A.M.}}:
\bbtitle{Quantum Engineering: Theory and Design of Quantum Coherent Structures}.
\bpublisher{Cambridge University Press},
\blocation{Cambridge}
(\byear{2011})
\end{bbook}
\endbibitem

\bibitem{VanDuzer:1999}
\begin{bbook}
\bauthor{\bsnm{Van~Duzer}, \binits{T.}},
\bauthor{\bsnm{Turner}, \binits{C.W.}}:
\bbtitle{Principles of Superconductive Devices and Circuits},
\bedition{2nd} edn.
\bpublisher{Prentice Hall},
\blocation{Upper Saddle River, NJ}
(\byear{1999})
\end{bbook}
\endbibitem

\bibitem{Tinkham:1996}
\begin{bbook}
\bauthor{\bsnm{Tinkham}, \binits{M.}}:
\bbtitle{Introduction to Superconductivity},
\bedition{2}nd edn.
\bpublisher{McGraw-{H}ill},
\blocation{New York}
(\byear{1996})
\end{bbook}
\endbibitem

\bibitem{Chen:1980}
\begin{bbook}
\bauthor{\bsnm{Chen}, \binits{W.-K.}}:
\bbtitle{Active Network and Feedback Amplifier Theory}.
\bpublisher{McGraw-Hill},
\blocation{New York}
(\byear{1980})
\end{bbook}
\endbibitem

\bibitem{Chua:1969}
\begin{bbook}
\bauthor{\bsnm{Chua}, \binits{L.O.}}:
\bbtitle{Introduction to Nonlinear Network Theory}.
\bpublisher{McGraw-{H}ill},
\blocation{New York}
(\byear{1969})
\end{bbook}
\endbibitem

\bibitem{Chua:1987}
\begin{bbook}
\bauthor{\bsnm{Chua}, \binits{L.O.}},
\bauthor{\bsnm{Desoer}, \binits{C.A.}},
\bauthor{\bsnm{Kuh}, \binits{E.S.}}:
\bbtitle{Linear and Non-Linear Circuits}.
\bsertitle{McGraw-Hill series in electrical and computer engineering}.
\bpublisher{McGraw-Hill Ryerson, Limited},
\blocation{New York}
(\byear{1987})
\end{bbook}
\endbibitem

\bibitem{Chua:2003}
\begin{barticle}
\bauthor{\bsnm{Chua}, \binits{L.O.}}:
\batitle{Nonlinear circuit foundations for nanodevices}.
\bjtitle{Proceedings of the IEEE}
\bvolume{91}(\bissue{11}),
\bfpage{1830}--\blpage{1859}
(\byear{2003})
\end{barticle}
\endbibitem

\end{thebibliography}

\end{document}